\documentclass[twocolumn,pra,aps,showpacs,amsmath,amssymb]{revtex4}

\usepackage{graphicx,latexsym}
\usepackage{epsfig}

\begin{document}
\title{Heisenberg-limited eavesdropping on the continuous-variable\\
quantum cryptographic protocol with no basis switching is impossible}

\author{J. Sudjana, L. Magnin\footnote{On leave from Laboratoire de l'Informatique 
et du Parall\'elisme, Ecole Normale Sup\'erieure de Lyon, 69364 Lyon cedex 07, France.}, R. Garc\'\i a-Patr\'on, and N. J. Cerf}
\affiliation{ QuIC, Ecole Polytechnique, CP 165, 
Universit\'e Libre de Bruxelles, 1050 Brussels, Belgium}

\date{June 2007}
\begin{abstract}
The Gaussian quantum key distribution protocol based on coherent states and
heterodyne detection [Phys. Rev. Lett. 93, 170504 (2004)] 
has the advantage that no active random basis switching
is needed on the receiver's side. Its security is, however, not very
satisfyingly understood today because the bounds on the secret key rate 
that have been derived from Heisenberg relations are not attained 
by any known scheme. Here, 
we address the problem of the optimal Gaussian individual attack against
this protocol, and derive tight upper bounds on the information
accessible to an eavesdropper. The optical scheme achieving this
bound is also exhibited, which concludes the security analysis of this
protocol.
\end{abstract}

\maketitle
\section{Introduction}

Over the past few years, an important research effort has been devoted to
continuous-variable quantum key distribution (QKD) protocols, motivated by
the prospects of realizing high-rate cryptosystems relying on homodyne
detection instead of photon counting. These systems also have the advantage
that they are based on standard (low-cost) telecom optical components, 
circumventing the need for single-photon sources nor single-photon detectors. 
In particular, 
Gaussian QKD protocols have been extensively investigated first because
they are conceptually simpler, but also mainly because their security
can be rigorously assessed. The first proposed Gaussian QKD protocol 
used squeezed states of light, which are modulated in one or the other
quadrature ($x$ or $p$) by the emitter (Alice), and are measured via
homodyne detection by the receiver (Bob) \cite{Cerf01}. 
Although this protocol is a very natural continuous-variable counterpart
of the famous BB84 protocol, its main drawback is the need for
a source of squeezed light.

A second Gaussian QKD protocol was devised, in which Alice generates
coherent states (instead of squeezed states) which are then modulated
both in $x$ and $p$, while Bob still performs homodyne detection \cite{GG02}.
Dealing with coherent states of light (simply produced 
with a laser) instead of squeezed or single-photon states makes 
this protocol very practical. This protocol, 
supplemented with the technique of reverse reconciliation, was experimentally 
demonstrated in Ref.~\cite{Nature03},
where it was shown that its range can, in principle, be arbitrarily large.
Note that, in these two protocols, Bob randomly chooses to homodyning one quadrature, either $x$ or $p$. In the squeezed-state
protocol, Bob then needs to reject the instances where he measured 
the other quadrature than the one modulated by Alice (this operation
is called sifting), which results in a decrease of the key rate 
by a factor 2 \footnote{This factor may actually be reduced and tend to 1
by making an asymmetric choice between $x$ and $p$ provided that the key length
is sufficiently large.}.
In the coherent-state protocol, Alice simply forgets the quadrature 
that is not measured by Bob, which may look like a loss of efficiency.
A third Gaussian protocol was therefore proposed,
in which Alice still transmits doubly-modulated coherent states
but Bob performs heterodyne measurements, that is, he measures both 
quadratures $x$ and $p$ simultaneously \cite{WL04} (this possibility
was also suggested for postselection-based protocols in \cite{lor04}). 
At first sight, this seems to imply that the rate is doubled, 
since Bob then acquires a pair of quadratures ($x,p$). Actually, since
heterodyne measurement effects one additional unit of vacuum noise on the 
measured quadratures, the two quadratures received by Bob are noisier 
than the single quadrature in the homodyne-based protocol.
The net effect, however, is generally an increase of the key rate when 
the two quadratures are measured simultaneously \footnote{This advantage of the heterodyne-based coherent-state protocol over the homodyne-based coherent-state
protocol is always true for a noiseless line, as well as for a noisy line
in reverse reconciliation.}.

This third protocol thus exhibits two advantages, namely that (i) the
key rate is generally higher than for the homodyne-based coherent-state
protocol, and (ii) there is no need to choose a random quadrature 
(i.e., no active basis choice is needed) at Bob's side. 
However, in order to make any definite statement on the security 
of this protocol, it is necessary to put precise limits on the maximum
information accessible to an eavesdropper (Eve).
Surprisingly, although bounds on the optimal Gaussian individual attack
against this protocol had been derived in \cite{WL04}, it has remained
unknown until now whether these bounds can be attained or not 
by an explicit eavesdropping strategy.
These bounds were derived using similar techniques 
to those used for the other Gaussian protocols, namely by writing Heisenberg uncertainty relations. Since for the protocols based on homodyne detection, the corresponding Heisenberg bounds can be attained by use of 
an explicit transformation (the entangling cloner),
it is tempting to conclude that the same is true for the heterodyne-based
protocol. On the other hand, since no explicit scheme has been found 
to date that saturates these bounds, another possibility is that 
these are loose, and tighter bounds remain to be found.

In this paper, we revisit the security of this coherent-state 
heterodyne-based Gaussian protocol, and prove that the second above
option is indeed true. We seek for the optimal Gaussian individual attack
by expressing the most general symplectic transformation characterizing
Eve's action and maximizing the information acquired by her. Restricting
to symplectic transformations is actually sufficient given that 
Gaussian attacks are provably optimal among individual attacks \cite{GC04}.
We conclude that this optimal attack is less powerful than expected,
in the sense that we derive a tighter bound than that based on the
Heisenberg inequalities. We also exhibit optical schemes that 
precisely attain this bound, both in direct and reverse reconciliation.
Hence, the resulting lower bound on the secret key rate is higher 
than that based on the Heisenberg uncertainty relations, 
making the heterodyne-based protocol even more efficient 
than originally thought.

\section{Heisenberg-limited eavesdropping}

\begin{figure}[t]
\begin{center}
\includegraphics[width=8.5cm]{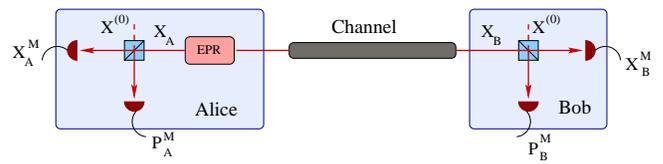}
\end{center}
\caption{Entanglement-based scheme of the protocol based on Alice sending coherent states and Bob applying heterodyne detection. 
Alice prepares an EPR state and
applies heterodyne detection on one half of it, resulting in
$(X_A^M,P_A^M)$, while the other half is sent to Bob. After transmission
via the channel, Bob performs an heterodyne measurement, resulting
in $(X_B^M,P_B^M)$. The superscript (0) indicates that the corresponding
state is the vacuum.}
\label{fig:CohHet}
\end{figure}

The Gaussian protocol based on coherent states and heterodyne detection \cite{WL04} can be shown to be equivalent to an entanglement-based 
scheme \cite{GC03}, where Alice prepares an EPR state and applies 
an heterodyne measurement on mode $A$, while Bob applies 
an heterodyne measurement on mode $B$. This is shown in Fig.~\ref{fig:CohHet}.
We restrict ourselves to individual attacks, where Eve completely 
controls the Alice-to-Bob channel separately for each transmitted state. 
Since Gaussian attacks are optimal among these attacks, we consider 
in what follows that Eve effects a Gaussian channel \footnote{Strictly 
speaking, the optimality proof of
Gaussian individual attacks given in Ref.~\cite{GC04} only applies
to DR protocols in which Alice sends squeezed states or RR protocols
in which Bob performs homodyne measurement. However, 
its extension to all Gaussian protocols, including the no-switching
protocol of interest here can be found in Ref.~\cite{RaulPhD}.}.
Consequently, the quantum state 
$\rho_{AB}$ before Alice and Bob's measurements can be assumed
to be a Gaussian two-mode state with a zero mean value and 
a covariance matrix $\gamma_{AB}$.
Usual Gaussian channels, such as optical fibers, effect a symmetric and
uncorrelated noise in both quadratures $x$ and $p$ 
(including, of course, the loss-induced noise),
so that we will only consider symmetric channels without $x$-$p$ correlations
in what follows. Since the EPR state (two-mode squeezed state) 
is also symmetric and exhibits no correlations between $x$ and $p$, 
we can write the resulting covariance matrix in a block-diagonal form as
\begin{equation}
\gamma_{AB}= 
\left(
\begin{array}{cc}
\gamma_{AB}^{x} & 0 \\
0& \gamma_{AB}^p
\end{array}
\right)	
\label{eq:EBCovMat}	
\end{equation}
with 
\begin{equation}
\gamma_{AB}^{x(p)}=
\left(
\begin{array}{cc}
V & \pm\sqrt{T(V^2-1)} \\
\pm\sqrt{T(V^2-1)} & T(V+\chi)
\end{array}
\right)
\end{equation}
where the signs $+$ and $-$ correspond to $\gamma_{AB}^{x}$ and
$\gamma_{AB}^{p}$, respectively. 
Here, $V$ is the variance of Alice's output thermal state, while
$T$ and $\chi=(1-T)/T+\epsilon$ are the transmittance and noise referred to the 
input of the Gaussian channel [the term $(1-T)/T$ stands for the loss-induced vacuum noise, while $\epsilon$ is the excess noise referred to the input].

In order to address the security of this protocol, we may, 
without loss of generality, assume that Eve holds the purification 
of the quantum state $\rho_{AB}$. By
measuring their systems, Bob and Eve then project Alice's share of the
joint pure state $|\Psi_{ABE}\rangle$ 
onto another pure state\footnote{We may indeed always assume
that Eve performs a measurement based on a {\it rank-one} 
Positive Operator Valued Measure (POVM), so that the resulting state
is pure. Otherwise, she would just need to disregard a part of her 
measuring system.}. Applying the Heisenberg
uncertainty relation on the pure state held by Alice (conditioning
on Bob and Eve's measurements), we have
\begin{equation}
V_{X_A|E}V_{P_A|B}\geq 1,
\label{eq:HeisenbergDR}
\end{equation}
where $X_A$ and $P_A$ are the canonically conjugate quadratures 
of Alice's mode and $V_{X|Y}$ is the conditional variance measuring
the remaining uncertainty on $X$ after the measurement of $Y$,
\begin{equation}
V_{X|Y}=\langle x^2\rangle-\frac{\langle xy\rangle^2}{\langle y^2\rangle},
\label{eq:CondVar}
\end{equation}
expressed in shot-noise units. 
Equation (\ref{eq:HeisenbergDR}) also has a symmetric counterpart that reads,
\begin{equation}
V_{P_A|E}V_{X_A|B}\geq 1.
\label{eq:HeisenbergDR2}
\end{equation}
Since we focus on a symmetric noise in $x$ and $p$, 
Eqs. (\ref{eq:HeisenbergDR}) and (\ref{eq:HeisenbergDR2}) 
can be unified into a single uncertainty relation
\begin{equation}
V_{A|E}V_{A|B}\geq 1.
\label{eq:HeisenbergDRf}
\end{equation}
where $A$ stands for any quadrature ($X_A$ or $P_A$) of Alice's mode.
This inequality will be used to put a lower bound on the uncertainty
of Eve's estimate of the key in Direct Reconciliation (DR), that is, 
when the key is made out of Alice's data while Bob and Eve compete 
to estimate it.
Similarly, in Reverse Reconciliation (RR), that is, when the key
is made out of Bob's data while Alice and Eve compete to estimate it,
one can derive a dual inequality 
\begin{equation}
V_{B|E}V_{B|A}\geq 1.
\label{eq:HeisenbergRRf}
\end{equation} 
where $B$ stands for any quadrature of Bob's mode. This will be used
to put a lower bound on the uncertainty of Eve's estimate of the key 
in RR.

Now, we will derive lower bounds on the secret key rates 
using the above uncertainty relations on the variances,
similarly as in Ref.~\cite{WL04}.
Restricting to individual attacks and one-way reconciliation, 
the DR and RR secret key rates for {\it each} of the two quadratures read 
\begin{eqnarray}
K^\text{DR}_{x\text{~or~}p}&=&H(A^M|E)-H(A^M|B^M),
\label{eq:KindDR2} \\
K^\text{RR}_{x\text{~or~}p}&=&H(B^M|E)-H(B^M|A^M),
\label{eq:KindRR2}
\end{eqnarray}
where $H(.)$ is the Shannon entropy, and $E$ stands for
Eve's optimal measurement maximizing her 
information (which is not necessarily the same in DR and RR).
Note that we use the variables $A^M$ and $B^M$ here (not $A$ and $B$),
since in this protocol Alice and Bob do not measure one single quadrature
but a pair of conjugate quadratures [$A^M$ ($B^M$) stands for the measurement
of one quadrature of mode $A$ ($B$), given that the conjugate quadrature is
simultaneously measured]. The total key rates $K^\text{DR}_{(x,p)}$ or
$K^\text{RR}_{(x,p)}$ derived later on are the sum of the
above expressions for $x$ and $p$.
If we assume that the channel is Gaussian, we can express the conditional 
entropies in Eqs.~(\ref{eq:KindDR2}) and (\ref{eq:KindRR2}) in terms of
conditional variances, so that the above Heisenberg
inequalities on conditional variances directly translate into bounds 
on the secret key rates.

\subsection{Direct reconciliation}

The problem of estimating Bob's uncertainty on Alice's measurements 
$A^M$ (that is, $X_A^M$ or $P_A^M$ knowing that the other one is 
also measured) can be reduced to estimating Bob's uncertainty 
on each of the quadratures of mode $A$ ($X_A,P_A$) since Alice's measurements 
result from mixing mode $A$ with vacuum on a balanced beam splitter,
see Fig.~\ref{fig:CohHet}.
Using Eqs. (\ref{eq:EBCovMat}) and (\ref{eq:CondVar}), one gets
\begin{equation}
V_{A|B}
=\frac{V\chi+1}{V+\chi}
\end{equation}
where $B$ stands for the same quadrature of mode $B$ ($X_B$ or $P_B$).
Similarly, using Eq.~(\ref{eq:CondVar}), and the fact that
$\langle (X_B^M)^2 \rangle = (1+ \langle (X_B)^2 \rangle)/2$
and $\langle X_A\, X_B^M \rangle = \langle X_A\, X_B \rangle)/\sqrt{2}$,
one gets
\begin{equation}
V_{A|B^M}
=\frac{T(V\chi+1)+V}{T(V+\chi)+1}
\end{equation}
which can then be converted into the variance of Bob's estimate 
of Alice's key
\begin{eqnarray}
V_{A^M|B^M}&=&\frac{1}{2}\Big[V_{A|B^M}+1\Big] \nonumber\\
&=&\frac{1}{2}\Big[\frac{(V+1)(T(\chi+1)+1)}{T(V+\chi)+1}\Big].
\end{eqnarray}
Using $V_{A|E}=1/V_{A|B}$ for the optimal eavesdropping
(since Bob {\it may} have performed homodyne detection and measured
one single quadrature), one gets for Eve's uncertainty 
on her estimate of Alice's key
\begin{align}
V_{A^M|E}&=\frac{1}{2}\Big[\frac{1}{V_{A|B}}+1\Big]  \nonumber \\
&=\frac{1}{2}\Big[\frac{(V+1)(\chi+1)}{V\chi+1}\Big]
\label{eq:varEveHeisenberg}
\end{align}
The secret key rate then reads,
\begin{align}
K^\text{DR}_{(x,p)}&=\log\Bigg[\frac{V_{A^M|E}}{V_{A^M|B^M}}\Bigg] \nonumber \\
&=\log\Bigg[\frac{(\chi+1)(T(V+\chi)+1)}{(V\chi+1)(T(\chi+1)+1)}\Bigg].
\label{eq:KDRCE}
\end{align}
Note that we have a factor two with respect to Eq.~(\ref{eq:KindDR2})
because the key is extracted from both quadratures $X_A^M$ and $P_A^M$.

\subsection{Reverse reconciliation}
Similarly, one can show that $V_{B|A}=T(\chi+1/V)$ and
$V_{B|A^M}=T(\chi+1)$, so that
the variance of Alice's estimate of Bob's data is
\begin{equation}
V_{B^M|A^M}=\frac{1}{2}\Big[V_{B|A^M}+1\Big]=\frac{1}{2}\Big[T(\chi+1)+1\Big].
\end{equation}
while, using $V_{B|E}=1/V_{B|A}$ (Alice {\it may} have performed
homodyne instead of heterodyne detection), one gets for
Eve's uncertainty
\begin{equation}
V_{B^M|E}=\frac{1}{2}\Big[\frac{1}{V_{B|A}}+1\Big]=
\frac{1}{2}\Big[\frac{T(V\chi+1)+V}{T(V\chi+1)}\Big]
\end{equation}
The secret key rate then reads,
\begin{align}
K^\text{RR}_{(x,p)}&=\log\Bigg[\frac{V_{B^M|E}}{V_{B^M|A^M}}\Bigg]
\nonumber \\
&=\log\Bigg[\frac{T(V\chi+1)+V}{T(V\chi+1)(T(\chi+1)+1)}\Bigg].
\label{eq:KRRCE}
\end{align}
We have a factor two with respect to Eq.~(\ref{eq:KindRR2})
because the key is extracted from both quadratures $X_B^M$ and $P_B^M$.

\section{Optimal Gaussian eavesdropping}

The entangling cloner, that is, the optimal attack against the homodyne-based  protocols \cite{GC03}, is clearly not optimal here as it 
allows to extract information about one single quadrature. We may think
of adapting it by applying an heterodyne detection on the mode that is
entangled with the mode injected in the line (as well as on the output mode
of Eve's beamsplitter simulating the losses). However, this is
equivalent to having a classical source of noise controlled by Eve, 
so that the optimal $V_{A(B)|E}$ that Eve can reach coincides with
the beamsplitter attack, which does not saturate (\ref{eq:KDRCE}) nor (\ref{eq:KRRCE}) as the excess noise $\epsilon$ only affects Alice and Bob mutual information but does not help Eve to reduce any uncertainty. 

Since the time when the heterodyne-based protocol was introduced \cite{WL04}, no attack has been found saturating bounds (\ref{eq:KDRCE}) and (\ref{eq:KRRCE}).
Logically, two possibilities remain open: (i) these bounds are tight 
but the optimal attacks reaching them remain to be found; (ii) these bound are not tight and the (unknown) optimal attacks can not saturate them. 
In order to answer this question, we need to search for the optimal attack
against this protocol with respect to all possible (individual Gaussian) 
attacks that Eve can do.
Although we are dealing with an infinite-dimensional Hilbert space, 
this task remains tractable because of the fact that Gaussian states
and operations have a simple characterization in terms of first- and
second-order moments of the quadratures. We thus need to find among
all possible linear canonical transformations the one which optimizes 
Eve's information either on Alice's data (DR) or on Bob's data (RR). 
Some symmetries also simplify the solution of this problem. 
Before searching for the optimal attack, 
let us consider these simplifications.

\subsubsection{Eve's Gaussian attack and the number of ancillae}
\begin{figure}[t]
\begin{center}
\includegraphics[width=8.5cm]{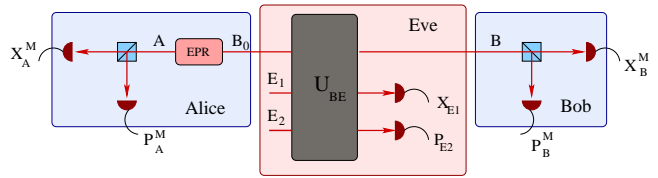}
\end{center}
\caption{Eve's attack against the protocol based on Alice sending coherent states and Bob applying heterodyne detection. 
Eve performs a unitary operation on her two ancillae $E_1$ and $E_2$
together with the mode $B_0$ sent by Alice.
She then measures $x$ on one ancilla and $p$ on the other one, in order to estimate simultaneously the two conjugate quadratures 
of Alice (DR) or Bob (RR).}
\label{fig:EveAttack}
\end{figure}

As we restrict Eve's attacks to Gaussian operations,
it is trivial to see that Eve must apply a Gaussian unitary transformation 
on the mode sent by Alice together with her ancillae, as shown in Fig.~\ref{fig:EveAttack}.
Indeed, applying a Gaussian completely positive maps instead 
of a unitary operation (i.e., discarding some ancillae) 
can only make Eve loose information on the secret key.
The number of ancillae that Eve needs is determined
as follows. First, it is easy to see that Eve needs at least two ancillary
modes to estimate either Alice's (DR) or Bob's (RR) quadratures, since one 
is needed to get $x$, the other to get $p$. Let us give an argument 
why these two ancillary modes are actually sufficient to implement the optimal 
attack. In the entanglement-based description, Eve holds 
the purification of $\rho_{AB}$, and therefore can be restricted to
occupy the same number of modes as $\rho_{AB}$, see \cite{HW01}. 
One should then be able to recover the entanglement-based scheme 
of Fig. \ref{fig:EveAttack} by applying a local unitary operation 
on Eve's side, since all purifications are equivalent up 
to a unitary operation on Eve's side. 

Thus, the optimal Gaussian attack we seek for corresponds,
in the Heisenberg picture, to a symplectic transformation $S$ acting
jointly on Alice's mode $B_0$ 
and Eve's ancillary modes $E_1$ and $E_2$, that is
\begin{align}
[\hat{x}_{B}, \hat{x}_{E_1}, \hat{x}_{E_2},& \hat{p}_{B}, \hat{p}_{E_1}, \hat{p}_{E_2}]^T = \nonumber \\
&S \; [\hat{x}_{B_0}, \hat{x}^{(0)}_{E_1}, \hat{x}^{(0)}_{E_2}, \hat{p}_{B_0}, \hat{p}^{(0)}_{E_1}, \hat{p}^{(0)}_{E_2}]^T,
\label{eq:Squadratures}
\end{align}
where the superscript $(0)$ is used to indicate that the corresponding
state is the vacuum. Then, Eve's optimal measurement 
on her two modes $E\equiv E_1E_2$ can be assumed to be a homodyne measurement
on these two modes in order to estimate either ($x_A,p_A$)
in DR or ($x_B,p_B$) in RR.

\subsubsection{Symmetric channel without $x$-$p$ correlations}


The symplectic transformation $S$ can be written
without loss of generality in a bloc-diagonal form as
\begin{equation}
S= 
\left(
\begin{array}{cc}
S_x & 0 \\
0 & S_p
\end{array}
\right)	
\label{eq:S}
\end{equation}
where $S_x$ and $S_p$ are related by the relation
\begin{equation}
S_p=(S_x^T)^{-1}
\label{eq:SCondition}
\end{equation}
in order to preserve the canonical commutation relations.
Indeed, we start with an initial Gaussian state of covariance matrix
$\gamma_{AB_0} \oplus \openone_{E_1 E_2}$, which is of the same form
as Eq.~(\ref{eq:EBCovMat}). More precisely, it is symmetric in $x$ and $p$
and admits no correlations between $x$ and $p$. After Eve's Gaussian
operation, we have a Gaussian state for modes $A$ and $B$, which,
by Schmidt decomposition, can be purified into a Gaussian 4-mode state
by extending the system with modes $E_1$ and $E_2$ \cite{HW01}. 
This can be understood
by applying a symplectic decomposition on modes $A$ and $B$
that converts their joint state into a product of two thermal states.
These thermal states can then be written as the reduction of EPR states, 
shared with Eve's modes $E_1$ and $E_2$. Since this symplectic decomposition
does not mix the $x$ and $p$ quadratures, the covariance matrix of the
4-mode pure state is again of the same form as Eq.~(\ref{eq:EBCovMat}).
Hence, the symplectic transformation $S$ applied by the eavesdropper
does not mix the $x$ and $p$ quadratures. We would like to stress
that this form, Eq.~(\ref{eq:S}),
is not an assumption but rather a simplification
originating from the fact that the channels of interest effect symmetric
uncorrelated noise in $x$ and $p$, as mentioned above.

The entry of the matrix $\gamma_{AB}^{x}$
corresponding to $\langle \hat{x}_B^2 \rangle = T(V+\chi)$
provides constraints on the first row of $S_x$, since we need to have
\begin{equation}
\hat{x}_B=\sqrt{T}(\hat{x}_{B_0} 
+\sqrt{\chi}\cos\theta \;\hat{x}^{(0)}_{E_1}
 +\sqrt{\chi}\sin\theta \; \hat{x}^{(0)}_{E_2}) 
\label{eq:x_B}
\end{equation}
where $\theta\in[0,2\pi]$ is a free parameter. Remember that
$\langle \hat{x}_{B_0}^2 \rangle =\langle \hat{x}_A^2 \rangle =V$.
Thus, we can write $S_x$ in general as
\begin{equation}
S_x=\sqrt{T}
\left(\begin{array}{ccc}
1 & \sqrt{\chi}\cos\theta & \sqrt{\chi}\sin\theta\\
a & b & c \\
r & s & t
\end{array}
\right)
\label{eq:Sx}
\end{equation}
where $\{a,b,c,r,s,t\}\in\mathbb{R}$ are six other free 
parameters. 
Using Equation~(\ref{eq:SCondition}), we can rewrite $S_p$ as
\begin{align}
&S_p=\frac{1}{d\sqrt{T}} \nonumber \\
&\times \left(\begin{array}{ccc}
bt-cs & cr-at & as-br\\
\sqrt{\chi}(s\sin\theta-t\cos\theta) & t-r\sqrt{\chi}\sin\theta & r\sqrt{\chi}\cos\theta-s \\
\underbrace{\sqrt{\chi}(c\cos\theta-b\sin\theta)}_{r'} & \underbrace{a\sqrt{\chi}\sin\theta-c}_{s'} 
& \underbrace{b-a\sqrt{\chi}\cos\theta}_{t'}
\end{array}
\right)
\label{eq:Sp}
\end{align}
where $d=\det (S_x)$. 
Given the symmetry of the channel,
the entry of $\gamma_{AB}^{p}$ corresponding to
$\langle \hat{p}_B^2 \rangle = T(V+\chi)$ provides a constraint 
on the first row of $S_p$, in a similar way as for $S_x$.
This yields the three conditions
\begin{eqnarray}
bt-cs&=&d\, T \nonumber \\
cr-at&=&d\, T\sqrt{\chi}\cos\phi \nonumber \\
as-br&=&d\, T\sqrt{\chi}\sin\phi.
\label{eq:SymCh}
\end{eqnarray}
where $\phi\in[0,2\pi]$ is a free parameter.
Finally, due to the symmetry of the channel in $x$ and $p$, we consider that 
Eve's optimal attack gives her the same uncertainty in $x$ and $p$.

\subsection{Direct reconciliation}

As before, Eve's uncertainty on Alice's measurements 
$A^M \equiv (X_A^M,P_A^M)$ can be calculated from 
the uncertainty of Eve on each of the two quadratures of mode 
$A$ ($X_A,P_A$). We have, 
for example, $V_{X_A^M|X_{E_1}}=\frac{1}{2}(V_{X_A|X_{E_1}}+1)$, 
and similarly for the $p$ quadrature.
The symmetry of Eve's information on $X_A$ and $P_A$ imposes that 
\begin{equation}
V_{X_A|X_{E_1}}=V_{P_A|P_{E_2}}\equiv V_{A|E}.
\label{eq:eqVae}
\end{equation}
Writing the second-order moments of $A$ and $E_1$,
\begin{eqnarray}
\langle \hat{x}_A^2\rangle&=&V \\
\langle \hat{x}_{E_1}^2\rangle&=&T(a^2V+b^2+c^2) \\
\langle \hat{x}_A\hat{x}_{E_1}\rangle&=&a\sqrt{T}\langle \hat{x}_A\hat{x}_{B_0}\rangle=a\sqrt{T(V^2-1)}
\end{eqnarray}
and plugging them into Eq.~(\ref{eq:CondVar}), we obtain
\begin{equation}
V_{X_A|X_{E1}}=\frac{V+\frac{a^2}{b^2+c^2}}{V\frac{a^2}{b^2+c^2}+1}.
\end{equation}
Similarly, one has for the $p$ quadrature
\begin{equation}
V_{P_A|P_{E2}}=\frac{V+\frac{r'^2}{s'^2+t'^2}}{V\frac{r'^2}{s'^2+t'^2}+1}.
\end{equation}
Finally, as a consequence of Eq.~(\ref{eq:eqVae}) we can write
\begin{equation}
V_{A|E}=\frac{V+\rho}{V\rho+1},
\label{eq:VaeRho}
\end{equation}
where 
\begin{equation}
\rho\equiv\frac{a^2}{b^2+c^2}=\frac{r'^2}{s'^2+t'^2}
\label{eq:defRho} 
\end{equation}
Given Eq.~(\ref{eq:x_B}), we see that
$\rho$ is proportional to the signal-to-noise ratio 
of the Alice-to-Eve channel (more precisely, 
the latter signal-to-noise ratio equals $\rho V$).
Thus, by definition, $\rho\geq 0$. Moreover, we can write
in analogy with Eq.~(\ref{eq:HeisenbergDR})
the Heisenberg uncertainty relation
\begin{equation}
V_{X_A|X_{E_1}} V_{P_A|P_{E_2}}\geq 1
\end{equation}
which, together with Eq.~(\ref{eq:eqVae}), 
implies that $V_{A|E}\geq 1$, or, equivalently, $\rho\leq 1$.
Note that the Heisenberg-limited attack in DR corresponds simply
to choose $\rho=\chi$.

We will now prove that such a choice is not possible, that is, it is
not consistent with the constraints we have on the matrices $S_x$ and $S_p$.
In order to further simplify $S_x$, we introduce the following 
change of variables,
\begin{eqnarray}
a&=&u\sqrt{\rho} \nonumber \\
b&=&u\sin\xi  \nonumber \\
c&=&u\cos\xi \label{eq:ChVar}
\end{eqnarray}
Using the variables $r',s',t'$ as defined in Eq.~(\ref{eq:Sp})
and the expression of $\rho$ in terms of these variables,
Eq.~(\ref{eq:defRho}), we then obtain
\begin{equation}
\Bigg(\frac{\chi-\rho}{\rho}\Bigg)\cos^2(\xi+\theta)=\Big(\sin(\xi+\theta)-\sqrt{\rho\chi}\Big)^2.
\label{eq:RhoCond1}
\end{equation}
Using the symmetry of the channel, Eq.~(\ref{eq:SymCh}), 
and the explicit expression of $d=\det S_x$, we obtain a second
similar equation
\begin{equation}
\Bigg(\frac{\chi-\rho}{\rho}\Bigg)\cos^2(\xi+\theta)=\Bigg(\sin(\xi+\theta)+\frac{1-T}{T\sqrt{\rho\chi}}\Bigg)^2,
\label{eq:RhoCond2}
\end{equation}
Expressing the equality 
between Eqs.~(\ref{eq:RhoCond1}) and (\ref{eq:RhoCond2}) 
yields two solutions. The first one, namely $\rho\chi=-(1-T)/T$, is unphysical
since $T\le 1$, $\rho\ge 0$, and $\chi\ge 0$. The second one yields
\begin{equation}
\sin(\xi+\theta)=\frac{1}{2}\frac{T\chi\rho-(1-T)}{T\sqrt{\chi\rho}}.
\label{eq:RhoSin}
\end{equation}
Furthermore, injecting Eq.~(\ref{eq:RhoSin}) into Eq.~(\ref{eq:RhoCond2}) gives
\begin{equation}
\cos^2(\xi+\theta)=\left( \frac{1}{2}\frac{T\chi\rho+(1-T)}{T\sqrt{\chi(\chi-\rho)}} \right)^2.
\label{eq:RhoCos}
\end{equation}
Finally, the relation $\cos^2(\xi+\theta)+\sin^2(\xi+\theta)=1$ 
provides us with a second-order equation in $\rho$,
\begin{equation}
T(T\chi^2+4)\rho^2-2\chi T(T+1)\rho+(1-T)^2=0
\label{eq:Rho2d}
\end{equation}
which always admits two solutions for a given channel (i.e. given parameters $T$ and $\chi$),
\begin{equation}
\rho_{\pm}=\frac{\chi T(T+1)\pm 2\sqrt{T[(T\chi)^2-(1-T)^2]}}{T(T\chi^2+4)}.
\end{equation}

Looking at Eq.~(\ref{eq:VaeRho}), we see that minimizing $V_{A|E}$ 
is equivalent to maximizing $\rho$, that is,
choosing $\rho_{+}$. Thus, Eve's minimum uncertainty on Alice's measurement
reads,
\begin{equation}
V_{A^M|E}^\text{min}=\frac{1}{2}\big[V_{A|E}^\text{min}+1\big]=\frac{1}{2}\frac{(V+1)(\rho_++1)}{V\rho_++1}
\label{eq:new1}
\end{equation}
and the lower bound on the DR secret key rate reads
\begin{align}
K^\text{DR}&=\log\Bigg[\frac{V_{A^M|E}^\text{min}}{V_{A^M|B^M}}\Bigg]  \nonumber \\
&=\log\Bigg[\frac{(\rho_++1)(T(V+\chi)+1)}{(V\rho_++1)(T(\chi+1)+1)}\Bigg].
\label{KDR}
\end{align}
Interestingly, Eq.~(\ref{eq:new1}) is similar
to its counterpart for the Heisenberg-limited attack,
Eq.~(\ref{eq:varEveHeisenberg}), but with $\rho_+$ replacing $\chi$.
It can easily be checked that $\rho_+ < \chi$, so that
the highest possible signal-to-noise ratio of the Alice-to-Eve channel 
is strictly lower than the one deduced from Heisenberg uncertainty 
relations. Hence, Eve's optimal attack is less powerful 
than expected from Heisenberg relations.

\begin{figure}[t]
\begin{center}
\includegraphics[width=9cm]{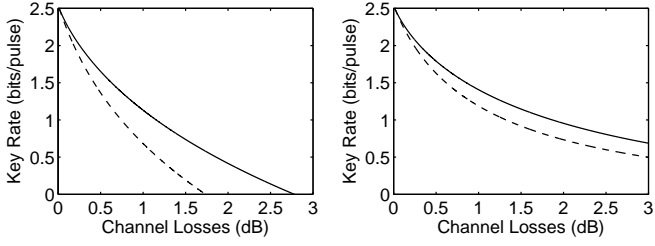}
\end{center}
\caption{Secret key rate as a function of the line losses 
for the optimal (solid line) and 
Heisenberg-limited (dashed line) attack. The curves are plotted for 
experimentally realistic values, $V=12$ and $\epsilon=0.01$,
in direct reconciliation (left panel)
or reverse reconciliation (right panel).
}
\label{Ch4:fig:HeisNew}
\end{figure}  

This is illustrated
in Fig.~\ref{Ch4:fig:HeisNew}, where the secret key rates
have been plotted for experimental realistic values 
of $V$ and $\epsilon$. The lower bound deduced from the Heisenberg
relations is satisfied, but loose with respect to the actual key rate.

\subsection{Reverse reconciliation}

Combining Eqs.~(\ref{eq:Squadratures}) and (\ref{eq:Sx}),
we obtain the second-order moments of $B$ and $E_1$
\begin{eqnarray}
\langle \hat{x}_B^2\rangle&=&T(V+\chi) \\
\langle \hat{x}_{E_1}^2\rangle&=&T(a^2V+b^2+c^2) \\
\langle \hat{x}_B \hat{x}_{E_1}\rangle&=&T(aV+b\sqrt{\chi}\cos\theta+c\sqrt{\chi}\sin\theta)
\end{eqnarray}
This results in  

\begin{widetext}
\begin{equation}
V_{X_B|X_{E1}}=T\frac{\Big[\frac{b^2+c^2}{a^2}+\chi-\frac{2\sqrt{\chi}}{a}(b\cos\theta+c\sin\theta)\Big]V+
\frac{\chi}{a^2}(b\sin\theta-c\cos\theta)^2}
{V+\frac{b^2+c^2}{a^2}}.
\end{equation}
where we have used Eq.~(\ref{eq:CondVar}).
Similarly, using the symmetry of the channel, Eq.~(\ref{eq:SymCh}),
we can write, 
\begin{equation}
V_{P_B|P_{E2}}=T\frac{\Big[\frac{s'^2+t'^2}{r'^2}+\chi-\frac{2\sqrt{\chi}}{r'}(s'\cos\phi+t'\sin\phi)\Big]V
+\frac{\chi}{r'^2}(s'\sin\phi-t'\cos\phi)^2}
{V+\frac{s'^2+t'^2}{r'^2}}
\end{equation}
\end{widetext}

Imposing the symmetry of Eve's information on $X_B$ and $P_B$
in analogy with Eq.~(\ref{eq:eqVae}), that is,
\begin{equation}
V_{X_B|X_{E1}}=V_{P_B|P_{E2}}\equiv V_{B|E},
\end{equation}
gives the three conditions
\begin{align}
\frac{r'^2}{s'^2+t'^2}&=\frac{a^2}{b^2+c^2}=\rho \label{eq:RhoCondRR}\\
\frac{s'\cos\phi+t'\sin\phi}{r'}&=\frac{b\cos\theta+c\sin\theta}{a}
=\frac{\sin(\xi+\theta)}{\sqrt{\rho}} \\
\frac{s'\sin\phi-t'\cos\phi}{r'}&=\frac{b\sin\theta-c\cos\theta}{a}
=\frac{\cos(\xi+\theta)}{\sqrt{\rho}} 
\end{align}
Note that condition (\ref{eq:RhoCondRR}) is exactly the same as in 
direct reconciliation. Surprisingly, it so happens that this condition
is sufficient to find an expression for $V_{B|E}$ which is the same 
as in direct reconciliation, making it unnecessary to use the other
two conditions. Indeed,
Eve's uncertainty on the quadratures of mode $B$ can be rewritten as
\begin{equation}
V_{B|E}=T\frac{\big[1+\chi\rho-2\sqrt{\chi\rho}\sin(\xi+\theta)\big]V+\chi\cos^2(\xi+\theta)}{V\rho+1}.
\end{equation}
Then, using the definition of $\sin(\xi+\theta)$  
coming from Eq.~(\ref{eq:RhoSin}) as well as Eq.~(\ref{eq:Rho2d}), 
we obtain
\begin{align}
&\cos^2(\xi+\theta)=\frac{\rho}{T\chi} \\
&1+\chi\rho-2\sqrt{\chi\rho}\sin(\xi+\theta)=1/T
\end{align}
which gives $V_{B|E}=V_{A|E}$. 
Therefore, just like in direct reconciliation, Eve's uncertainty
on the quadratures of mode $B$ is minimized by choosing $\rho_+$,
\begin{equation}
V_{B|E}^\text{min}=\frac{V+\rho_+}{V\rho_++1}.
\end{equation}
Then, Eve's uncertainty on Bob's measured values becomes
\begin{equation}
V_{B^M|E}^\text{min}=\frac{1}{2}\Big[V_{B|E}^\text{min}+1\Big]=\frac{1}{2}\frac{(V+1)(\rho_++1)}{V\rho_++1},
\end{equation}
so that the RR secret key rate reads
\begin{align}
K^\text{RR}&=\log\Bigg[\frac{V_{B^M|E}^\text{min}}{V_{B^M|A^M}}\Bigg]  \nonumber \\
&=\log\Bigg[\frac{(V+1)(\rho_++1)}{(V\rho_++1)(T(\chi+1)+1)}\Bigg].
\label{KRR}
\end{align}
This rate is illustrated in Fig.~\ref{Ch4:fig:HeisNew}, where it is
compared with the lower bound deduced from the Heisenberg
relations in RR. We conclude again that the Heisenberg-limited
attack is not reachable.

\begin{figure}[t]
\begin{center}
\includegraphics[width=9cm]{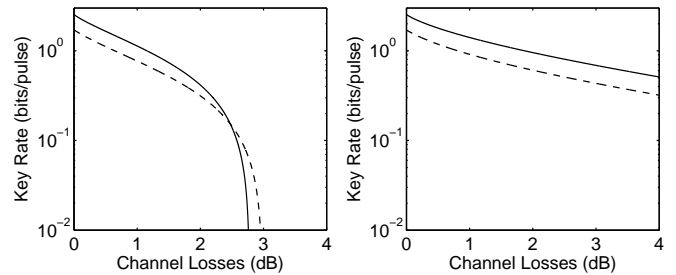}
\end{center}
\caption{Secret key rate as a function of the line losses  
for the heterodyne-based (solid line) and homodyne-based (dashed line)
protocols in direct reconciliation (left panel)
or reverse reconciliation (right panel).
We use experimentally realistic values, 
$V=12$ and $\epsilon=0.01$, and consider that Alice sends coherent states
in both cases.}
\label{Ch4:fig:All2}
\end{figure}

For illustration, we compare in Fig.~\ref{Ch4:fig:All2} 
the secret key rate of the coherent-state {\it homodyne-based} protocol
to that of the present coherent-state {\it heterodyne-based}
protocol in direct and reverse reconciliation 
[Eqs.~(\ref{KDR}) and (\ref{KRR})]. For realistic parameters 
$V$ and $\epsilon$, we notice that the heterodyne-based protocol 
always yields higher rates than the homodyne-based protocol in RR. 
This also means that the maximum tolerable excess noise $\epsilon$ 
in RR is higher with the heterodyne-based protocol 
regardless the losses. In DR, the heterodyne-based protocol
gives an advantage over the homodyne-based protocol only for line losses
below some threshold. This threshold can be shown to decrease 
for increasing $\epsilon$, so that the maximum tolerable noise 
is actually higher for the homodyne-based protocol in DR.

\section{Optical setup achieving the best Gaussian attack}

In Section~III, we have reduced the problem of maximizing Eve's information
to that of optimizing a single parameter $\rho$, 
the other parameters remaining free. This implies that the optical implementation of the best Gaussian attack is not unique. In this Section,
we present two particularly interesting examples of such an optical implementation, namely the teleportation attack 
and the ``feed-forward'' attack. Note that the latter attack was
also considered in Ref.~\cite{WL04}, where it was noticed that it
curiously does not reach the Heisenberg limit.

\subsection{Teleportation attack}

\begin{figure}[t]
\begin{center}
\includegraphics[width=8.5cm]{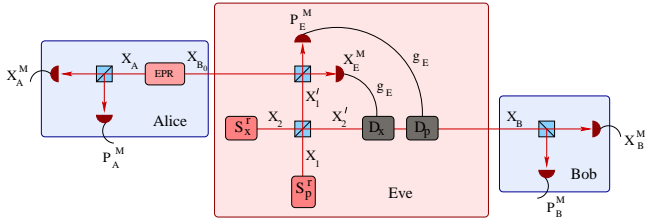}
\end{center}
\caption{Teleportation attack against the (entanglement-based scheme of the)  Gaussian protocol based on Alice sending coherent states and Bob applying heterodyne detection. Eve first generates an EPR pair ($E'_1,E'_2$) by mixing a  $x$-squeezed vacuum state ($E_2$) with a $p$-squeezed vacuum state ($E_1$) at a balanced beamsplitter. Then, she performs a Bell measurement on
Alice's outgoing mode $B_0$ together with $E'_1$. 
Depending on the measurement outcome and the fixed gain $g_E$,
she then displaces mode $E'_2$ by $x$ ($D_x$) and $p$ ($D_p$).
The resulting state is sent to Bob. By tuning the squeezing parameter
$r$ and the gain $g_E$, Eve can simulate any
Gaussian channel ($T,\chi$) and extract the optimal amount of information.}
\label{fig:TelAtt}
\end{figure}

The teleportation attack consists in Eve applying a continuous-variable 
quantum teleportation where the input is Alice's outgoing mode 
and the output is given to Bob, as shown in Fig.~\ref{fig:TelAtt}. 
Eve extracts information from the outcomes ($X_E^M,P_E^M$)
of her Bell measurement performed on Alice's outgoing mode $B_0$ 
together with one of the modes ($E'_1$) of an EPR state.
It is easy to see that there are two limiting cases.
If the squeezing factor $r$ of the EPR pair is zero, 
implying that $E'_1$ is in a vacuum state, then the scheme becomes
equivalent to an heterodyne measurement of $B_0$ by
Eve followed by the classical preparation of a coherent state 
(the vacuum state in mode $E'_2$ which is displaced by some amount
depending on $X_E^M$ and $P_E^M$).
This situation corresponds to an entanglement-breaking channel 
giving no secret key. On the contrary, if the squeezing factor $r$ is infinite,
the teleportation succeeds perfectly and Eve gets no information at all due to the infinite noise in the thermal state $E'_1$. This situation corresponds to a perfect channel with no losses and no excess noise ($T=1,\epsilon=0$). 
We will now show that for any intermediate value of $r$,
such a teleportation attack can be made optimal.

Since all the involved canonical transformations are symmetric in $x$ and $p$, 
we will detail the proof for the $x$ quadrature only. Eve starts by
preparing two squeezed vacuum states, one in mode $E_2$ (squeezed in $x$) and  the other is mode $E_1$ (squeezed in $p$),
\begin{eqnarray}
\hat{x}_1&=&e^{r}\hat{x}_1^{(0)} \\
\hat{x}_2&=&e^{-r}\hat{x}_2^{(0)},
\end{eqnarray}
and mixes them on a balanced beamsplitter, thereby generating an EPR state
\begin{eqnarray}
\hat{x}'_1&=&[e^{r}\hat{x}_1^{(0)}+e^{-r}\hat{x}_2^{(0)}]/\sqrt{2} \\
\hat{x}'_2&=&[e^{r}\hat{x}_1^{(0)}-e^{-r}\hat{x}_2^{(0)}]/\sqrt{2}.
\end{eqnarray}
Eve then applies a Bell measurement by mixing $E'_1$ and $B_0$ on a balanced beamsplitter, and measuring $x$ on one output and $p$ on the other,
\begin{equation}
\hat{x}_{E^M}=\frac{1}{\sqrt{2}}[\hat{x}_{B_0}+\hat{x}'_1]=\frac{1}{\sqrt{2}}\hat{x}_{B_0}+\frac{1}{2}[e^r\hat{x}_1^{(0)}+e^{-r}\hat{x}_2^{(0)}].
\end{equation}
Next, Eve displaces her mode $E'_2$ 
by an amount proportional to the measurement outcome $X_E^M$
(multiplied by the classical gain $g_E$) and sends it to Bob, giving
\begin{align}
\hat{x}_B&=\hat{x}'_2+g_E\, \hat{x}_{E^M}   \nonumber \\
&=\frac{g_E}{\sqrt{2}}\hat{x}_{B_0}+\frac{e^r}{\sqrt{2}}\Big[1+\frac{g_E}{\sqrt{2}}\Big]\hat{x}_1^{(0)}
+\frac{e^{-r}}{\sqrt{2}}\Big[1-\frac{g_E}{\sqrt{2}}\Big]\hat{x}_2^{(0)}.
\end{align}
In order to comply with $\langle \hat{x}_B^2\rangle=T(V+\chi)$,
 we need to fix $g_E$ and $r$ in such a way that
\begin{eqnarray}
g_E&=&\sqrt{2T} \\
T\chi&=&(1+T)\cosh2r+2\sqrt{T}\sinh2r. \label{eq:TelCond}
\end{eqnarray}
 
\subsubsection{Direct reconciliation.}
Writing the second-order moments of $\hat{x}_A$ and $\hat{x}_{E}$, namely
\begin{eqnarray}
\langle\hat{x}_A^2\rangle&=&V \\
\langle\hat{x}_{E}^2\rangle&=&(V+\cosh2r)/2 \label{eq:TelEveVar}\\
\langle\hat{x}_{A}\hat{x}_{E}\rangle&=&\langle\hat{x_A}\hat{x}_{B_0}\rangle/\sqrt{2}
=\sqrt{(V^2-1)/2}
\end{eqnarray}
one can show, using Eq.~(\ref{eq:CondVar}), that Eve's uncertainty 
on Alice's data is
\begin{equation}
V_{A|E}=\frac{V\cosh2r+1}{V+\cosh2r}.
\end{equation}
By choosing 
\begin{equation}
\rho=\frac{1}{\cosh2r}
\label{eq:Optr}
\end{equation}
this expression for $V_{A|E}$ coincides with Eq.~(\ref{eq:VaeRho}).
Combining Eq.~(\ref{eq:TelCond}) with the relation 
$\cosh^22r-\sinh^22r=1$, we see that $\rho$ must satisfy the second-order polynomial equation (\ref{eq:Rho2d}), whose solution
gives the value of $\rho$ that optimizes Eve's information.
Equation~(\ref{eq:Rho2d}) having two possible solutions $\rho_{\pm}$
generating the same quantum channel ($T,\chi$), we then have two possible
solutions for the squeezing parameter $r$.
Looking at Eq.~(\ref{eq:Optr}), we see that that the squeezing parameter corresponding to the optimal choice $\rho_+$ is the lowest of the
two solutions since it corresponds 
to the minimum added noise on Eve's measurement. 

\subsubsection{Reverse reconciliation.}
Using Eqs.~(\ref{eq:CondVar}), (\ref{eq:TelCond}), (\ref{eq:TelEveVar}),
and
\begin{equation}
\langle\hat{x}_B\hat{x}_E\rangle=\frac{1}{\sqrt{2}}\big[V\sqrt{T}+\sinh 2r+\sqrt{T}\cosh 2r\big],
\end{equation}
one can show that Eve's uncertainty on each of Bob's quadratures reads
\begin{equation}
V_{B|E}=\frac{V\cosh2r+1}{V+\cosh2r}=V_{A|E},
\end{equation}
implying that the teleportation attack is also optimal (choosing the lowest
squeezing parameter) for the reverse reconciliation protocol.

\subsection{Feed-forward attack}

\begin{figure}[t]
\begin{center}
\includegraphics[width=8.5cm]{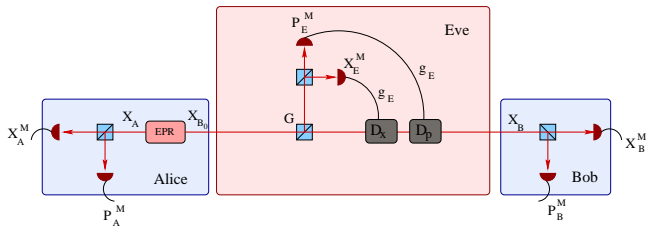}
\end{center}
\caption{Entanglement based scheme of Eve ``feed-forward'' attack over the protocol based on Alice sending coherent states and Bob applying heterodyne detection. Eve extract part of the signal sent by Alice using a beamsplitter
(transmittance $G$) and applies en heterodyne detection on it.
Depending on the measurement result times a given fixed gain $g_E$ Eve 
displaces mode $E'_2$ over $x$ ($D_x$) and $p$ ($D_p$).
The resulting state is then sent to Bob.
By tuning the transmittance of the beamsplitter ($G$) and the gain ($g_E$) Eve can simulate any
Gaussian channel ($T,\chi$) and extract the optimal amount of information.}
\label{fig:FFAtt}
\end{figure}

In the case of a noisy channels with no losses ($T=1$) 
and direct reconciliation, Eve's optimal teleportation attack 
is exactly the same scheme as the one proposed in Ref.~\cite{AF06} to 
reach an optimal tradeoff between disturbance and state estimation
for coherent states (when the success of both processes is measured
using the fidelity). This is not surprising since optimally estimating 
the coherent state sent by Alice while minimizing its disturbance 
is exactly what Eve attempts to achieve in her optimal attack in direct reconciliation.
In Ref.~\cite{AF06}, two alternative schemes to the teleportation reaching the same optimal tradeoff were also presented, the ``feed-forward'' attack 
and the asymmetric cloning machine. 
Those two schemes can very naturally be extended to our case ($T\le 1$) 
if we allow for different mean values for the input and output modes, 
which gives rise to new optical schemes for the optimal attack.

For example, it can be checked that Eve can realize an optimal attack
(both in DR and RR) using the ``feed-forward'' scheme described 
in Fig.~\ref{fig:FFAtt}
by fixing the parameters of the beamsplitter transmittance $G$ and the feed-forward gain $g_E$ as
\begin{eqnarray}
G&=&\frac{1-\rho_+}{1+\rho_+} \\
g_E&=&\big(\sqrt{T}-\sqrt{G}\big)\sqrt{\frac{2}{1-G}}.
\end{eqnarray}

\section{Conclusion}

We have revisited the security of the Gaussian quantum cryptographic protocol
with no basis switching (with Alice sending coherent states and Bob
performing heterodyne measurements) introduced in Ref.~\cite{WL04}.
We have considered the most general Gaussian individual attack against
this protocol by characterizing an arbitrary symplectic transformation
and maximizing Eve's information over all such transformations. 
We have found that, in contrast with all other Gaussian protocols 
that had been studied so far, no attack exists that attains the security 
bounds deduced from the Heisenberg uncertainty relations, 
making these bounds unreachable in the present case. A tight bound was derived,
both in direct and reverse reconciliation, and several explicit optical 
schemes that attain this bound have been exhibited. Remarkably, 
this makes the coherent-state heterodyne-based Gaussian protocol better
than what was implicitly assumed in the original analysis~\cite{WL04}.

We may wonder what is so special about this no-switching protocol?
As a matter of fact, in the two Gaussian protocols 
based on homodyne detection, one of the two quadratures plays a special role,
namely the one that is measured by Bob (provided, in the squeezed-state
protocol, that it is also the one modulated by Alice; otherwise the instance
is discarded). The Heisenberg uncertainty relations then express 
that any action on this quadrature, which carries the key, translates into
some additional noise on the dual quadrature. Monitoring the noise on this
dual quadrature then puts an upper limit on the information potentially 
acquired by Eve on the key-carrying quadrature. This simple and very intuitive 
interpretation fails for the heterodyne-based protocol because then 
both quadratures must be treated together (Alice modulates both quadratures
and Bob measures both quadratures). The security can be viewed as 
resulting from kind of an information conservation law 
through a ``fan-out'' channel (leading to both Bob and Eve),
akin to what is observed in the optimal estimation-vs-disturbance tradeoff 
for coherent states \cite{AF06} or in the asymmetric Gaussian cloning
of coherent states \cite{FC07}.

We acknowledge financial support from the EU under projects
COVAQIAL and SECOQC, and from the IUAP programme of the Belgian government 
under the project {\tt PHOTONICS@BE}. R.G.-P. acknowledges support from the Belgian foundation FRIA.

{\it Note added}: The findings of this paper have also been obtained
simultaneously and independently in \cite{unpublished}.


\end{document}